\documentclass[fleqn,twoside]{article}
\usepackage{espcrc2}

\usepackage{graphicx}

\newcommand{\AmS}{{\protect\the\textfont2
  A\kern-.1667em\lower.5ex\hbox{M}\kern-.125emS}}

\title{Astrophysical sources of high energy neutrinos}

\author{E. Waxman\address[WIS]{Physics Faculty, Weizmann Institute of
        Science, Rehovot 76100, Israel\\ waxman@wicc.wiezmann.ac.il}
        \thanks{To appear in Nuclear Physics B (Proceedings Supplement),
        Proc. XXth International Conference on
        Neutrino Physics and Astrophysics (Munich 2002)}}

\begin{document}

\begin{abstract}

Several high energy, $>100$~GeV, neutrino telescopes are currently
operating or under construction. Their main motivation is the
extension of the horizon of neutrino astronomy to cosmological
scales. We show that general, model independent, arguments imply
that $\sim1$~Gton detectors are required to detect cosmic high
energy neutrino sources. Predictions of models of some of the
leading candidate sources, gamma-ray bursts and micro-quasars, are
discussed, and the question of what can be learned from neutrino
observations is addressed.

\vspace{1pc}
\end{abstract}

\maketitle

\section{Introduction and summary}
\label{sec:summary}

High energy, $>100$~GeV, neutrino telescopes are currently
operating in deep lake water (BAIKAL,
http://baikal-neutrino.da.ru/) and under Antarctic ice (AMANDA,
http://amanda.berkeley.edu/amanda/). Two under-water detectors are
currently under construction in the Mediterranean (ANTARES,
http://antares.in2p3.fr/; NESTOR, http://www.cc.uoa.gr/~nestor/),
aiming at achieving effective volumes $\sim0.1{\rm km^3}$,
comparable to that of AMANDA. Much larger, $\simeq1{\rm km^3}$
telescopes are under construction in Antarctic ice (the IceCube
extension of AMANDA, http://icecube.wisc.edu/), and under
development in the Mediterranean (NEMO,
http://nemoweb.lns.infn.it/). For a detailed review of the
experiments, and also of their scientific goals, see \cite{HENAP}.

The driving motivation for the construction of km-scale neutrino
telescopes is the observation of cosmic point sources. At present,
several neutrino telescopes monitor solar MeV neutrinos
\cite{SolarNuTalks}, and may also detect MeV neutrinos from
supernova explosions, such as supernova 1987A, in our local
galactic neighborhood. The construction of high-energy neutrino
telescopes is aimed at extending the distances accessible to
neutrino astronomy to cosmological scales. This new window onto
the cosmos will provide a probe of the most powerful sources in
the universe through observations of high-energy neutrinos.

The existence of extra-Galactic high energy neutrino sources is
implied by the observations of ultra-high energy (UHE),
$>10^{19}$~eV, cosmic-rays. The cosmic-ray spectrum flattens at
$\sim10^{19}$~eV \cite{fly,agasa}. There are indications that the
spectral change is correlated with a change in composition, from
heavy to light nuclei~\cite{fly,composition}. These
characteristics, which are supported by analysis of Fly's Eye,
AGASA and HiRes-MIA data, and for which some evidence existed in
previous experiments~\cite{Watson91}, suggest that the cosmic ray
flux is dominated at energies $< 10^{19}$~eV by a Galactic
component of heavy nuclei, and at UHE by an extra-Galactic source
of protons. Also, both the AGASA and Fly's Eye experiments report
an enhancement of the cosmic-ray flux near the Galactic disk at
energies $\le10^{18.5}$~eV,  but not at higher
energies~\cite{anisotropy}.

Irrespective of the nature of the cosmic-ray sources, some
fraction of these particles will produce pions as they escape from
the acceleration site, either through hadronic collisions with
ambient gas or through interaction with ambient photons, leading
to  electron and muon neutrino production from the decay of
charged pions.

In \S~\ref{sec:nu-bound} we discuss the upper bound implied by UHE
cosmic-ray observations on the diffuse high energy neutrino flux.
In \S~\ref{sec:km-scale} we show that general, model independent
arguments imply that detectors with masses equal to or larger than
1~Gton, equivalent to 1~km$^3$ of water, should be constructed in
order to detect the expected neutrino signal in the energy range
of 1 to $10^3$~TeV. This conclusion applies both for the detection
of point sources and for the detection of diffuse extra-Galactic
flux. Even larger detectors may be required at higher energies.
The construction of detectors with effective volume $\gg1~{\rm
km}^3$ at energies $\gg10^3$~TeV may require the use of techniques
different than the under-water/ice optical Cerenkov technique
employed by the AMANDA, ANTARES, Baikal, IceCube, NEMO, and NESTOR
experiments. Such alternative technique may be, e.g., the
detection of coherent radio emission from $>10^{16}$~eV neutrino
induced cascades \cite{alternatives}.

The origin of UHE cosmic-rays is one of the most exciting open
questions of high energy astrophysics \cite{SiglNaganoWatson}. The
extreme energy of the highest energy events poses a challenge to
models of particle acceleration. Very few known astrophysical
objects have characteristics indicating that they may allow
acceleration of particles to the observed high energies. The
detection of high energy neutrinos may resolve the puzzle of UHE
cosmic-ray origin, as the high energy neutrinos, unlike the
charged cosmic-ray protons which are subject to deflection by
magnetic fields, will point directly to their sources.

The most powerful known extra-galactic objects, gamma-ray bursts
(GRBs) and active galactic nuclei (AGN), are candidate sources for
the production of UHE cosmic-rays, and are therefore likely
sources of neutrinos in the energy range of 1 to $10^3$~TeV. GRBs
are transient flashes of $\sim1$~MeV gamma-rays lasting typically
for 1 to 100~s, that are observed from sources at cosmological
distances. The apparent isotropic luminosity of GRBs is
$\sim10^{52}$~erg/s. They are believed to be powered by the rapid
accretion of a fraction of a solar mass of matter onto a newly
born solar-mass black hole. AGN are persistent sources with
apparent luminosity reaching $\sim10^{48}$~erg/s. They are thought
to be powered by mass accretion onto $10^{6}$--$10^9$ solar-mass
black holes, that reside at the centers of galaxies. In both GRBs
and AGN, mass accretion is believed to drive a relativistic plasma
outflow that results in the acceleration of high-energy particles,
which emit non-thermal radiation. A similar process could also
power Galactic micro-quasars, which may be considered as a
scaled-down version of AGN, powered by stellar-mass black holes or
neutron stars.

Despite the overall success of models in explaining the observed
phenomena associated with these high energy sources, the models
are largely phenomenological and our understanding of underlying
physical processes is still incomplete. In all cases, neutrino
observations will provide unique information on the physics of the
underlying engine.

In \S~\ref{sec:GRBs} we discuss neutrino emission from GRBs. We
describe the underlying model and its predictions for neutrino
telescopes, and address the question of what can be learned from
neutrino observations. The discussion illustrates how neutrino
observations will help resolving open questions related to the
physics that underlies models of high energy astrophysical
sources. In \S~\ref{sec:MQs} we briefly review neutrino emission
from micro-quasars.

Finally, it should be emphasized that a detection of even a
handful of neutrino events correlated with GRBs will allow to test
for neutrino properties, e.g. flavor oscillation and coupling to
gravity, and to place constraints on deviation from Lorentz
invariance with accuracy many orders of magnitude better than
currently possible. We discuss this in some more detail in
\S~\ref{sec:basic-physics}.

\section{An upper bound to the diffuse neutrino flux}
\label{sec:nu-bound}

We first derive in \S~\ref{sec:CR-rate} the rate of generation of
UHE protons implied by cosmic-ray observations. We then derive in
\S~\ref{sec:bound} the upper bound on the diffuse neutrino flux.

\subsection{The UHE cosmic-ray generation rate}
\label{sec:CR-rate}

Fly's Eye stereo spectrum is well fitted in the energy range
$10^{17.6}$~eV to $10^{19.6}$~eV by a sum of two power laws: A
steeper component, with differential number spectrum $J\propto
E^{-3.50}$, dominating at lower energy, and a shallower component,
$J\propto E^{-2.61}$, dominating at higher energy, $E>10^{19}$~eV.
The data are consistent with the steeper component being composed
of heavy nuclei primaries, and the lighter one being composed of
proton primaries. The observed UHE cosmic-ray flux and spectrum
may be accounted for by a two component, Galactic + extra-Galactic
model \cite{BnW02}. For the Galactic component, this model adopts
the Fly's Eye fit,
\begin{equation}
\frac{dn}{dE} ~\propto~ E^{-3.50}. \label{eq:galacticspectrum}
\end{equation}
The extra-Galactic proton component is derived in this model by
assuming that extra-galactic protons in the energy range of
$10^{19}$~eV to $10^{21}$~eV are produced by
cosmologically-distributed sources at a rate
\begin{equation}
\dot{\varepsilon}^{\rm CR}\approx 3\times 10^{44} {\rm
erg~Mpc^{-3}~yr^{-1}}, \label{eq:energyrate}
\end{equation}
with a power law differential energy spectrum
\begin{equation}
\frac{dn_p}{dE_p} \propto E_p^{-n}\,,\quad n\approx 2.
\label{eq:energyspectrum}
\end{equation}
The corresponding energy per logarithmic decade of protons is
\begin{equation}
E_p^2\frac{d\dot{n}_p^{\rm CR}}{dE_p}\approx0.7\times 10^{44} {\rm
erg~Mpc^{-3}~yr^{-1}}. \label{eq:GRB_p_rate}
\end{equation}
The spectral index, $n\approx2$, is that expected for acceleration
in sub-relativistic collisionless shocks in general, and in
particular for the GRB model discussed in \S~\ref{sec:GRB-model}.
The energy generation rate, Eq.~\ref{eq:energyrate}, is motivated
by the GRB energy generation rate, compare Eq.~\ref{eq:GRB_p_rate}
with Eq.~\ref{eq:GRB_e_rate}.

\begin{figure}
\includegraphics[width=3.3in]{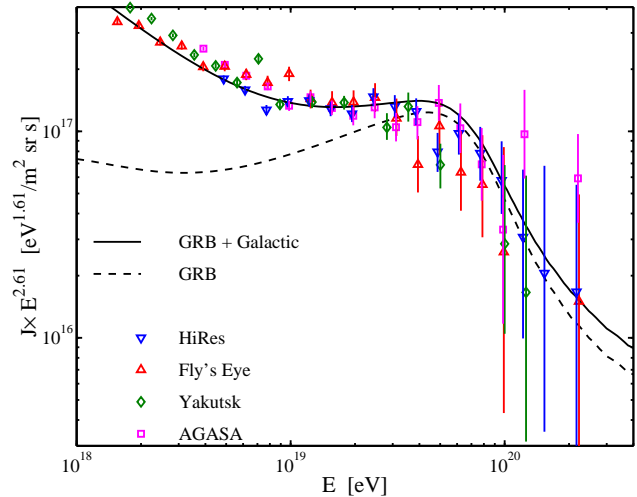} \caption{Model versus data. The solid curve shows
the energy spectrum derived from the two-component model discussed
in Section~\ref{sec:CR-rate}. The dashed curve shows the
extra-Galactic proton contribution, Eq.~\ref{eq:GRB_p_rate}.}
\label{fig:data_model}
\end{figure}

Figure~\ref{fig:data_model} compares the model prediction with the
data from the AGASA~\cite{agasa}, Fly's Eye~\cite{fly},
Hires~\cite{HiRes}, and Yakutsk~\cite{yakutsk} cosmic ray
experiments\footnote{The Haverah Park data have recently been
re-analyzed using modern numerical simulations of air-shower
development~\cite{HP}. The reanalysis resulted in significant
changes of inferred cosmic-ray energies compared to previously
published results (\cite{Watson91} and references quoted therein).
This improved analysis is available only at energies
$<10^{19}$~eV.}. The absolute flux measured at $10^{19}$~eV
differs between the various experiments, corresponding to a
systematic $\simeq11\%$ ($\simeq19\%$, $\simeq-7.5\%$)
over-estimate of event energies in the AGASA (Yakutsk, HiRes)
experiments compared to the Fly's Eye experiment (see also
\cite{Yoshida95,BnW02}). These systematic shifts are well within
the systematic uncertainties quoted by the experiments. We have
therefore applied the appropriate small systematic shifts in
absolute energy, to bring the various experiments into agreement.
The results are not sensitive to the choice of absolute energy
scale \cite{BnW02}.

The choice of cosmological model is unimportant for cosmic ray
energies above $10^{19}$ eV, which is the region of interest here.
For definitness, a flat universe with $\Omega_m = 0.3$ and
$\Omega_{\Lambda} = 0.7$, and Hubble constant $H_0=65{\,\rm
km/s\,Mpc}$ was assumed in Fig.~\ref{fig:data_model}. Also,
Eq.~\ref{eq:energyrate} represents the local, $z=0$, energy
generation rate. Motivated by the evidence for association of GRBs
with star formation, it was assumed that the generation rate
evolves with redshift $z$ like $f(z)=(1+z)^3$ at low redshift,
$z<1.9$, $f(z)={\rm Const.}$ for $1.9<z<2.7$, and an exponential
decay at $z>2.7$ \cite{SFR}.

The model described above is similar to that proposed in
\cite{W95a}. The improved constraints on UHECR spectrum and flux
provided by the recent observations of HiRes introduce only a
small change, compared to \cite{W95a}, to the inferred energy
generation rate, given by Eq.~\ref{eq:energyrate}.

The suppression of model flux above $10^{19.7}$~eV is due to
energy loss of high energy protons in interaction with the
microwave background, i.e. to the ``GZK cutoff'' \cite{GZK}. The
spectra measured by Fly's Eye, HiRes and Yakutsk are consistent
with the model, and hence with the existence of a GZK cutoff.

The data from the AGASA experiment, the exposure of which is
$\sim1/3$ of the combined exposure of the other experiments, is
consistent with the model, and with the other experiments, up to
$10^{20}$~eV, but show an excess of events above $10^{20}$~eV. The
origin of this discrepancy is not yet clear. However, it should be
pointed out that since the $>10^{20}{\rm eV}$ flux is dominated by
sources at distances $\le 50\ {\rm Mpc}$, over which the
distribution of known astrophysical systems (e.g. galaxies,
clusters of galaxies) is inhomogeneous, significant deviations
from model predictions presented in Fig.~\ref{fig:data_model} for
a uniform source distribution are expected at this energy
\cite{W95a}. Clustering of cosmic-ray sources leads to a standard
deviation, $\sigma$, in the expected number, $N$, of events above
$10^{20}$ eV, given by $\sigma /N = 0.9(d_0/10 {\rm Mpc})^{0.9}$
\cite{clustering}, where $d_0$ is the unknown scale length of the
source correlation function and $d_0\sim10$ Mpc for field
galaxies.

\subsection{The flux bound}
\label{sec:bound}

The energy generation rate given by Eq.~\ref{eq:GRB_p_rate} sets
an upper limit to the neutrino flux that may be produced by
sources which, like GRBs and the observed jets of AGN, are
optically thin to pion producing interactions of protons with
source photons (or ambient nucleons) \cite{WBbound}.

If the high-energy protons produced by the extra-galactic sources
lose a fraction $\epsilon<1$ of their energy through production of
pions before escaping the source, the resulting present-day energy
density of muon neutrinos is $E_\nu^2 dN_\nu/ dE_\nu
\approx0.25\epsilon t_H E_{p}^2d\dot n_{p}/dE_{p}$, where
$t_H\approx10^{10}{\rm yr}$ is the Hubble time. For energy
independent $\epsilon$, the neutrino spectrum follows the proton
generation spectrum, since the fraction of the proton energy
carried by a neutrino produced through a photo-meson interaction,
$E_\nu\approx0.05E_p$, is independent of the proton energy. The
$0.25$ factor arises because neutral pions, which do not produce
neutrinos, are produced with roughly equal probability with
charged pions, and because in the decay of charged pions muon
neutrinos carry approximately half the charged pion energy. Thus,
an upper limit to the muon neutrino flux ($\nu_\mu$ and
$\bar\nu_\mu$ combined) is obtained for $\epsilon=1$
\cite{WBbound},
\begin{eqnarray}
E_\nu^2 \Phi_\nu &&\le E_\nu^2 \Phi_\nu^{\rm WB}
\approx0.25\xi_Zt_H{c\over4\pi}E_{p}^2{d\dot n_{p}^{\rm CR}\over
dE_{p}} \cr && \approx1.5\times10^{-8}\xi_Z\frac{\rm GeV}{\rm
cm^{2}\,s\,sr}\,. \label{eq:nu-bound}
\end{eqnarray}

In the derivation of Eq.~\ref{eq:nu-bound} we have neglected the
redshift energy loss of neutrinos produced at cosmic time $t<t_H$,
and implicitly assumed that the cosmic-ray generation rate per
unit (comoving) volume is independent of cosmic time. The quantity
$\xi_Z$ in Eq.~\ref{eq:nu-bound} has been introduce to describe
corrections due to redshift evolution and energy loss. Assuming
that the UHE proton, and hence neutrino, energy generation rate
evolves rapidly with redshift, following the evolution of star
formation rate \cite{SFR}, see \S~\ref{sec:CR-rate}, we find that
$\xi_Z\approx3$ (with weak dependence on cosmology). The
correction is small despite the strong evolution with redshift,
since the universe spends only a small fraction of its present age
at high $z$. For no evolution, we have $\xi_Z\approx0.6<1$ (with
weak dependence on cosmology) due to redshift energy loss of
neutrinos.

There are two speculative types of sources for which the
Waxman-Bahcall bound, Eq.~\ref{eq:nu-bound}, does not apply. These
sources, for which we have no observational evidence to date,
could in principle produce a neutrino flux exceeding the
Waxman-Bahcall limit. The first special type of source is one in
which neutrinos are produced by processes other than photo-meson
or proton-nucleon interactions. Such sources include, e.g., decays
of topological defects or of supermassive dark matter particles
(see, e.g., \cite{SiglNaganoWatson}). The second type of special
source is one for which the optical depth for photo-production of
mesons (or for proton-nucleon interaction) is high. Examples of
such optically thick scenarios include neutrinos produced in the
cores of AGNs (rather than in the jets), or in a collapsing
galactic nucleus \cite{thick}.

\section{Phenomenological considerations: km$^3$-scale detectors}
\label{sec:km-scale}

\subsection{Detection of point sources}
\label{sec:point}

One may obtain an estimate for the required telescope size by
considering the minimal flux of a source that may be detected by a
neutrino telescope of effective area $A$ (in the plane
perpendicular to the source direction) and exposure time $T$. The
probability that a muon, produced by interaction of a muon
neutrino with a nucleon, will cross the detector is given by the
ratio of the muon and neutrino mean free paths, which (for water,
ice) is approximately given by
$P_{\mu\nu}=10^{-4}(E_\nu/100{\rm\,TeV})^\alpha$, with $\alpha=1$
for $E_\nu<100$~TeV and $\alpha=0.5$ for $E_\nu>100$~TeV. A source
of energy flux $f_\nu$ in neutrinos of energy $E_\nu$ will produce
$N=(f_\nu/E_\nu)P_{\mu\nu}AT$ events in the detector. Thus, the
flux required for the detection of $N$ events is
\begin{eqnarray}
f_\nu\sim10^{-11}N&&\left(\frac{E_\nu}{100{\rm\,TeV}}\right)^{1-\alpha}
\cr\times &&\left(\frac{AT}{\rm km^2 yr}\right)^{-1}\,\frac{\rm
erg}{\rm cm^2s}. \label{eq:fmin}
\end{eqnarray}
For cosmological sources, with
characteristic distance of $d\sim c/H0\sim4$~Gpc, the minimum
luminosity of a detectable source is therefore
\begin{equation}
L_\nu>10^{47}\left(\frac{E_\nu}{100{\rm\,TeV}}\right)^{1-\alpha}
\left(\frac{AT}{\rm km^2 yr}\right)^{-1}\,{\rm erg/s}.
\label{eq:Lmin}
\end{equation}

\begin{figure}
\includegraphics[width=2.95in]{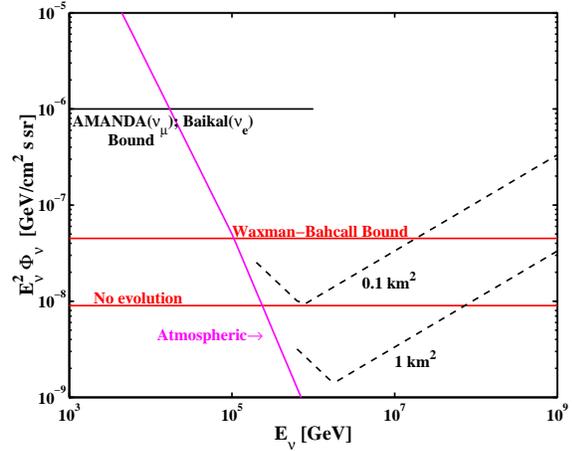} \caption{The projected sensitivity of
km-scale detectors, based on Eqs.~\ref{eq:Phimin} and
\ref{eq:Phibgnd}, compared with the Waxman-Bahcall upper bound on
the diffuse neutrino intensity, Eq.~\ref{eq:nu-bound}. Also shown
are the current experimental upper bounds imposed by the Baikal
and AMANDA experiments \cite{exp-bounds}, and the atmospheric
neutrino background.} \label{fig:bound}
\end{figure}
Objects with luminosity $\ge10^{47}{\rm erg/s}$, more than 13
orders of magnitude higher than the Solar luminosity, are very
rare. The only known class of steady sources that produce such
high luminosity are AGN. It is therefore clear that km-scale
neutrino detectors are required for the detection of cosmological
sources. This argument holds also for transient sources: The
brightest known transients are GRBs, which produce luminosity
$\sim10^{52}{\rm erg/s}$ over $\sim100$~s duration. Replacing
$T=1$~yr with $T=100$~s in the above equation, implies a minimum
luminosity $L_\nu\sim10^{52}{\rm erg/s}$.

A lower limit to the source flux is also set by the requirement
that the signal would exceed the background produced by
atmospheric neutrinos. The atmospheric neutrino flux is
approximately given by
$10^{-8.5}(E_\nu/500{\rm\,TeV})^{-\beta}{\rm GeV/cm^2s\,sr}$, with
$\beta=1.7$ for $E_\nu<500$~TeV and $\beta=2$ for $E_\nu>500$~TeV.
Given that the expected angular resolution of the telescopes is
approximately $\theta=1$~degree, the source flux for which the
signal constitutes a $5\sigma$ detection over the atmospheric
background flux is
\begin{eqnarray}
f_\nu\sim10^{-12}&&\left(\frac{E_\nu}{10^{14.5}{\rm
eV}}\right)^{-0.8} \cr&&\times\frac{\theta}{1\rm
deg}\left(\frac{AT}{\rm km^2 yr}\right)^{-1/2}\,\frac{\rm erg}{\rm
cm^2s}. \label{eq:fbgnd}
\end{eqnarray}
Thus, for km-scale detectors the atmospheric neutrino background
does not pose more stringent constraints on source flux than those
imposed by the requirement for a detectable signal (except at low
energies).

\subsection{Detection of diffuse background}
\label{sec:diffuse}

Using arguments similar to those used in \S~\ref{sec:point}, we
may obtain the minimum intensity of a diffuse neutrino background,
that will allow its detection by a detector of effective area $A$
observing $2\pi$~sr of the sky over a duration $T$. The intensity
required for the detection of $N$ events is
\begin{eqnarray}
E_\nu^2\Phi_\nu\sim10^{-9}N&&\left(\frac{E_\nu}{100{\rm\,TeV}}\right)^{1-\alpha}
\cr&&\times\left(\frac{AT}{\rm km^2 yr}\right)^{-1}\,\frac{\rm
GeV}{\rm cm^2s\,sr}, \label{eq:Phimin}
\end{eqnarray}
and the minimum intensity required for a $5\sigma$ detection over
the atmospheric background is
\begin{eqnarray}
E_\nu^2\Phi_\nu\sim10^{-8}&&\left(\frac{E_\nu}{10^{14.5}{\rm\,eV}}\right)^{-0.8}
\cr&&\times\left(\frac{AT}{\rm km^2 yr}\right)^{-1/2}\,\frac{\rm
GeV}{\rm cm^2s\,sr}. \label{eq:Phibgnd}
\end{eqnarray}

We compare in Figure~\ref{fig:bound} the projected sensitivity of
km-scale detectors, based on Eqs.~\ref{eq:Phimin} and
\ref{eq:Phibgnd}, with the upper bound Eq.~\ref{eq:nu-bound} on
the diffuse intensity, and with the current experimental upper
bounds imposed by the Baikal and AMANDA experiments
\cite{exp-bounds}. Similar to the case of point sources, km-scale
detectors are required to achieve the sensitivity that may allow
detection of a diffuse background. Note that, as mention in
\S~\ref{sec:summary}, at energy $E_\nu\gg10^3$~TeV the required
effective volume is $\gg1{\rm km}^3$.

\section{GRBs: candidate extra-Galactic sources of $>1$~TeV neutrinos}
\label{sec:GRBs}

\subsection{GRB model and UHE proton production}
\label{sec:GRB-model}

General phenomenological considerations, based on $\gamma$-ray
observations, indicate that, regardless of the nature of the
underlying sources, GRBs are produced by the dissipation of the
kinetic energy of a relativistic expanding fireball
\cite{AG_th_review}. A compact source, $r_0\sim10^7$~cm, produces
a wind, characterized by an average luminosity $L\sim10^{52}{\rm
erg\ s}^{-1}$ and mass loss rate $\dot M$. At small radius, the
wind bulk Lorentz factor, $\Gamma$, grows linearly with radius,
until most of the wind energy is converted to kinetic energy and
$\Gamma$ saturates at $\Gamma\sim L/\dot M c^2\sim300$.
Variability of the source on a time scale $\Delta t\sim10$~ms,
resulting in fluctuations in the wind bulk Lorentz factor $\Gamma$
on a similar time scale, results in internal shocks in the ejecta
at a radius $r\sim r_d\approx\Gamma^2c\Delta t\gg r_0$. It is
assumed that internal shocks reconvert a substantial part of the
kinetic energy to internal energy, which is then radiated as
$\gamma$-rays by synchrotron and inverse-Compton radiation of
shock-accelerated electrons. At a later stage, the shock wave
driven into the surrounding medium by the expanding fireball
ejecta leads to the emission of lower-energy "afterglow."

Since the observed radiation is produced at a large distance, $\gg
r_0$, from the underlying source, the model is largely independent
of the nature of the underlying compact object, as long as this
object is capable of producing a wind with the properties implied
by observations. The underlying GRB progenitors are yet unknown.
Collapse of massive stars and mergers of compact objects (e.g. of
binary neutron stars) are the most widely discussed candidates
\cite{AG_th_review}.

The observed radiation is produced, both during the GRB and the
afterglow, by synchrotron emission of shock accelerated electrons.
In the region where electrons are accelerated, protons are also
expected to be shock accelerated. The internal shocks within the
expanding wind are mildly relativistic (in the wind rest frame),
and hence we expect our understanding of non-relativistic shock
acceleration to apply to the acceleration of protons in these
shocks. In particular, the predicted energy distribution of
accelerated protons is expected to be $dn_p/dE_p\propto E_p^{-2}$
\cite{Blandford87}, similar to the predicted electron energy
spectrum, which is consistent with the observed photon spectrum.

Several constraints must be satisfied by wind parameters in order
to allow proton acceleration to high energy $E_p$. We summarize
below these constraints. The reader is referred to
\cite{W95,W01rev,W02} for a more detailed discussion. The
requirement that the acceleration time be smaller than the wind
expansion time sets a lower limit to the strength of the wind
magnetic field, which may be expressed as a lower limit to the
ratio of magnetic field to electron energy density \cite{W95},
\begin{equation}
u_B/u_e>0.02 \Gamma_{2.5}^2 E_{p,20}^2L_{\gamma,52}^{-1},
\label{eq:xiB}
\end{equation}
where $E_p=10^{20}E_{p,20}$~eV, $\Gamma=10^{2.5}\Gamma_{2.5}$ and
$L_{\gamma}=10^{52}L_{\gamma,52}{\rm erg/s}$ is the wind
$\gamma$-ray luminosity. A second constraint is imposed by the
requirement that the proton acceleration time be smaller than the
proton energy loss time, which is  dominated by synchrotron
emission. This sets an upper limit to the magnetic field strength,
which in turn sets a lower limit to $\Gamma$ \cite{W95}
\begin{equation}
\Gamma>130 E_{p,20}^{3/4}\Delta t^{-1/4}_{-2}. \label{eq:Gmin}
\end{equation}
Here, $\Delta t=10^{-2}\Delta t_{-2}$~s. As explained in
\cite{W95}, the constraints Eq.~\ref{eq:xiB} and Eq.~\ref{eq:Gmin}
hold regardless of whether the fireball is a sphere or a narrow
jet (as long as the jet opening angle is $>1/\Gamma$). The
luminosity in Eq.~\ref{eq:xiB} is the "isotropic equivalent
luminosity", i.e. the luminosity under the assumption of isotropic
emission.

Internal shocks within the wind occur over a wide range of radii,
corresponding to a wide range of variability time scales: $\Delta
t\sim1$~ms, the dynamical time of the source, to $\Delta
t\sim1$~s, the wind duration. Protons are therefore accelerated to
high energy over a wide range of radii. In particular, at large
radii the external medium affects fireball evolution, and a
"reverse shock" is driven backward into the fireball ejecta and
decelerates it, due to the interaction with the surrounding
medium. This shock is also mildly relativistic, and its parameters
are similar to those of an internal shock with $\Delta t\sim10$~s.
Protons may therefore be accelerated to ultra-high energy in this
shock as well \cite{WnB-AG,W01rev}.

The constraints Eq.~\ref{eq:xiB} and Eq.~\ref{eq:Gmin} are
remarkably similar to those inferred from $\gamma$-ray
observations: $\Gamma>300$ is implied by the $\gamma$-ray spectrum
in order to avoid high pair-production optical depth, magnetic
field close to equipartition, $u_B/u_e\sim0.1$, is required in
order to account for both $\gamma$-ray emission and afterglow
observations \cite{AG_th_review}. This suggests that GRBs and UHE
s may originate from common sources.

The suggested association between GRBs and the sources of UHE
protons is further strengthened by comparing the proton energy
generation rate, Eq.~\ref{eq:energyrate}, with the GRB energy
generation rate. The evidence for association of GRB sources with
star-formation \cite{AG_th_review}, suggests that the GRB rate
evolves with redshift following the star-formation rate. Under
this assumption, the local ($z=0$) GRB rate is $\approx
5\times{\,10^{-10}\rm Mpc^{-3}~yr^{-1}}$ \cite{Schmidt01} and
their average 0.1~MeV to 2~MeV $\gamma$-ray energy release is
$\approx 2.5\times10^{53}$~erg \cite{Frail01}, corresponding to an
energy generation rate of\footnote{The quoted GRB rate density and
$\gamma$-ray energy assume that the GRB emission is isotropic. If
emission is confined to a solid angle $\Delta\Omega<4\pi$, then
the GRB rate is increased by a factor $(\Delta\Omega/4\pi)^{-1}$
and the GRB energy is decreased by the same factor. However, their
product, the energy generation rate, is independent of the solid
angle of emission. It is currently believed that GRB's are beamed
on average into a solid angle of $4\pi/500$ \cite{Frail01}, which
implies a rate higher (total energy lower) by a factor of 500
compared to values inferred assuming isotropic emission.}
\begin{equation}
\dot{\varepsilon}_\gamma^{\rm GRB}\approx 1.3\times 10^{44} {\rm
erg~Mpc^{-3}~yr^{-1}}. \label{eq:GRB_g_rate}
\end{equation}
This rate is remarkably similar to the energy generation rate of
UHE protons inferred from cosmic-ray observations,
Eq.~\ref{eq:energyrate}. Note, that although the proton generation
rate, Eq.~\ref{eq:energyrate}, is approximately twice the GRB
$\gamma$-ray generation rate, Eq.~\ref{eq:GRB_g_rate}, the
corresponding energy per logarithmic decade of electrons implied
by Eq.~\ref{eq:GRB_g_rate},
\begin{equation}
E_e^2\frac{d\dot{n}_e^{\rm GRB}}{d\varepsilon_e}\approx 10^{44}
{\rm erg~Mpc^{-3}~yr^{-1}}. \label{eq:GRB_e_rate}
\end{equation}
is similar to Eq.~\ref{eq:GRB_p_rate}. Thus, GRBs may be the
sources of observed UHE cosmic-rays, provided they produce similar
energy in MeV $\gamma$-rays, or, equivalently in high energy
electrons, and in high energy protons.

\subsection{$\sim100$~TeV neutrinos}
\label{sec:generic-nu}

Protons accelerated in the fireball to high energy lose energy
through photo-meson interaction with fireball photons. The decay
of charged pions produced in this interaction results in the
production of high energy neutrinos. The key relation is between
the observed photon energy, $E_\gamma$, and the accelerated
proton's energy, $E_p$, at the threshold of the
$\Delta$-resonance. In the observer frame,
\begin{equation}
E_\gamma \,E_{p} = 0.2 \, {\rm GeV^2} \, \Gamma^2\,.
\label{eq:keyrelation}
\end{equation}
For $\Gamma\approx300$ and $E_\gamma=1$~MeV, we see that
characteristic proton energies $\sim 10^{16}$~eV are required to
produce pions. Since neutrinos produced by pion decay typically
carry $5\%$ of the proton energy, production of $\sim 10^{14}$~eV
neutrinos is expected \cite{WnB97,WBbound}.

The fraction of energy lost by protons to pions, $f_\pi$ is
\cite{WnB97,GSW01,W01rev}
\begin{eqnarray}
f_\pi(\epsilon_p)\approx0.2{L_{\gamma,52}^{1/3} \Delta
t_{-2}^{-1/3}}.\label{eq:fpi}
\end{eqnarray}
Assuming that GRBs generate similar energy in high-energy protons
and electrons, accounting therefore for the observed UHE
cosmic-ray flux, then using Eq.~\ref{eq:nu-bound}, the expected
GRB muon-neutrino flux is
\begin{eqnarray}
E_\nu^2\Phi_{\nu} &&\approx 0.2
\frac{f_\pi}{0.2}E_\nu^2\Phi_{\nu}^{\rm WB} \cr&&\approx
0.9\times10^{-8}{f_\pi\over0.2}{\rm GeV\,cm}^{-2}{\rm s}^{-1}{\rm
sr}^{-1}. \label{eq:JGRB}
\end{eqnarray}
This neutrino spectrum extends to $\sim10^{16}$~eV, and suppressed
at higher energy due to energy loss of pions and muons. Comparing
Eq.~\ref{eq:JGRB} with Eq.~\ref{eq:Phibgnd}, we find that $\sim20$
neutrino-induced muon events per year are expected (over
$4\pi$~sr) in a cubic-km detector. Note, that GRB neutrino events
are correlated both in time and in direction with gamma-rays, and
hence there detection is practically background free \cite{WnB97}.

\subsection{$\sim10^{17}$~eV "Afterglow" neutrinos}
\label{sec:AG-nu}

High energy neutrino emission may also result from photo-meson
interactions of protons accelerated to high energies in the
reverse shocks driven into the fireball ejecta at the initial
stage of interaction of the fireball with its surrounding gas,
which occurs on time scale $T\sim10$~s (see
\S~\ref{sec:GRB-model}). Optical--UV photons are radiated by
electrons accelerated in shocks propagating backward into the
ejecta. The interaction of these low energy, 10~eV--1~keV, photons
and high energy protons produces a burst of duration $\sim T$ of
ultra-high energy, $10^{17}$--$10^{19}$~eV, neutrinos, as
indicated by Eq.~\ref{eq:keyrelation} \cite{WnB-AG}.

The expected muon neutrino intensity depends on the density of the
surrounding medium \cite{WnB-AG,Dai00}, which may differ by orders
of magnitude between models assuming different GRB progenitors. If
GRB fireballs expand into typical density inter-stellar medium,
$n\sim1{\rm cm}^{-3}$, as expected if the GRBs are the result of
mergers of compact objects, the expected intensity is
\begin{eqnarray}
E_\nu^2\Phi_\nu\approx 10^{-10} \left({E_\nu\over10^{17}{\rm
eV}}\right)^{\beta} {\rm GeV\,cm}^{-2}{\rm s}^{-1}{\rm sr}^{-1},
\label{eq:JGRBAG}
\end{eqnarray}
with $\beta=1/2$ for $E_\nu>10^{17}{\rm eV}$ and $\beta=1$
otherwise. If GRB fireballs expand into massive stellar winds
($n\sim10^4{\rm cm}^{-3}$), as expected if GRBs result from the
collapse of massive stars, then
\begin{equation}
E_\nu^2\Phi_\nu\approx 10^{-8}\min\{1, E_\nu/10^{17}{\rm eV}\}
\frac{\rm GeV}{\rm cm^{-2}s\,sr}. \label{eq:JGRBAGw}
\end{equation}
The neutrino flux is expected to be strongly suppressed at energy
$E_\nu>10^{19}$~eV, since protons are not expected to be
accelerated to energy $E_p\gg10^{20}$~eV.

\subsection{$\sim5$~TeV "Collapsar" neutrinos}
\label{sec:collapsar-nu}

Lower energy, $\sim5$~TeV, neutrino emission may be expected in
the case where GRBs originate from the collapse of massive stars
\cite{MnW01}. In this scenario, accretion onto a black hole,
produced by the collapse of the massive star (the "collapsar")
core, drives a relativistic jet that propagates through the
stellar envelope along the collapsar rotation axis. The shocks
producing the $\gamma$-rays must occur after the fireball has
emerged from the stellar envelope. While the jet is making its way
out of the star, its rate of advance is slowed down in a
termination shock that heats the stellar plasma to keV
temperatures, and additional internal shocks are expected in the
pre-deceleration interior jet. The latter can accelerate protons
to $>10^5$ GeV, which interact with the X-ray photons in the
stellar jet cavity leading to pion production and hence electron
and muon neutrinos (and anti neutrinos) with energies $E_{\nu}
\ge5$~TeV. These neutrinos appear as a precursor signal, lasting
for time scales of tens of seconds prior to the observation of any
$\gamma$-rays produced outside the star by a collapsar-induced
GRB. The TeV neutrino fluence from an individual collapse at
cosmological distance $z\sim 1$ implies $\sim0.1$ upward moving
muons per collapse in a 1~km$^3$ detector.

\subsection{Implications: GRB models}
\label{sec:implications}

The emission of $\sim100$~TeV neutrinos discussed in
\S~\ref{sec:generic-nu} is independent of the underlying GRB
progenitor. It is a natural outcome of the "generic" fireball
model described in \S~\ref{sec:GRB-model}, where the observed
$\gamma$-rays are produced by internal shocks within a dissipative
relativistic wind. The major underlying assumption, upon which the
predictions depend, is that the fireball momentum is carried
(after the initial stage of acceleration) by relativistic protons.
The predicted flux, Eq.~\ref{eq:JGRB}, implies $\sim20$
neutrino-induced muon events per year in a cubic-km detector.
Neutrino telescopes may therefore allow to test underlying
assumptions of the fireball model. Detection of the predicted
signal will also provide strong support for the model of UHE
cosmic-ray production in GRBs.

The emission of $\sim10^{17}$~eV neutrinos, \S~\ref{sec:AG-nu}, is
also a natural prediction of the fireball model. The expected
neutrino intensity depends strongly, however, on the GRB
progenitor, compare Eqs.~\ref{eq:JGRBAG} and~\ref{eq:JGRBAGw}.
Detection of such UHE neutrinos may therefore provide constraints
on the progenitor type.

Finally, emission of $\sim5$~TeV "collapsar" neutrinos is expected
only for collapsar GRB progenitors. The detection of a $\sim5$~TeV
neutrino precursor will therefore provide a clear signature of the
collapsar model.

\subsection{Implications: Lorentz invariance, The weak equivalence principle,
and neutrino oscillations}
\label{sec:basic-physics}

Detection of neutrinos from GRBs could be used to test the
simultaneity of neutrino and photon arrival to an accuracy of
$\sim1$~s ($\sim1$~ms for short bursts), checking the assumption
of special relativity that photons and neutrinos have the same
limiting speed \cite{WnB97}. These observations would also test
the weak equivalence principle, according to which photons and
neutrinos should suffer the same time delay as they pass through a
gravitational potential. With 1~s accuracy, a burst at a distance
of 1~Gpc would reveal a fractional difference in limiting speed
$\sim10^{-17}$, and a fractional difference in gravitational time
delay of order $10^{-6}$ (considering the Galactic potential
alone). Previous applications of these ideas to supernova 1987A,
where simultaneity could be checked only to an accuracy of order
several hours, yielded much weaker upper limits: of order
$10^{-8}$ and $10^{-2}$ for fractional differences in the limiting
speed and time delay, respectively.

The model discussed above predicts the production of high energy
muon and electron neutrinos with a 2:1 ratio (tau neutrinos may be
produced by photo-production of charmed mesons; however, the
higher energy threshold and lower cross-section for charmed-meson
production, compared to pion production, typically imply that the
ratio of charmed-meson to pion production is $10^{-4}$
\cite{WnB97}).  Because of neutrino oscillations, neutrinos that
get here are expected to be almost equally distributed between
flavors for which the mixing is strong. Upgoing taus, rather than
muons, would be a distinctive signature of such oscillations. It
may be possible to distinguish between taus and muons in a
km$^3$-scale detector, since at $10^3$~TeV the tau decay length is
$\sim1$~km. This will allow a "tau appearance experiment".

\section{Micro-quasars: Galactic candidate sources}
\label{sec:MQs}

The jets associated with Galactic micro-quasars \cite{Mirabel99}
are believed to be ejected by accreting stellar-mass black holes
or neutron stars. Much like for AGN, the content of the jets is an
open issue. The dominant energy carrier in the jet is at present
unknown (with the exception of the jet in SS433). Scenarios
whereby energy extraction is associated with spin-down of a Kerr
(rotating) black hole favor electron-positron composition
(although baryon admixture is an issue), while scenarios in which
an initial rise of the X-ray flux leads to ejection of the inner
part of the accretion disk imply electron-positron jets, as widely
claimed to be suggested by the anti-correlation between the X-ray
and radio flares seen during major ejection events. A possible
diagnostic of electron-positron jets is the presence of
Doppler-shifted spectral lines, such as the H$\alpha$ lines as
seen in SS433.  The detection of such lines from jets having a
Lorentz-$\Gamma$ factor well in excess of unity (as is the case in
the super-luminal micro-quasars) may, however, be far more
difficult than in SS433, as the lines are anticipated to be very
broad.

Neutrino telescopes may prove to be a much sharper diagnostic
tool. If the energy content of the jets in the transient sources
is dominated by electron-proton plasma, then a several hour
outburst of 1 to 100~TeV neutrinos (and high-energy photons)
produced by photo-production of pions should precede the radio
flares associated with major ejection events \cite{LnW01}. Several
neutrinos may be detected during a single outburst by a
km$^3$-scale detector \cite{LnW01,diStefano02}, thereby providing
a powerful probe of micro-quasar jet physics and of their
innermost structure.


\begin{thebibliography}{9}

\bibitem{HENAP}
  HENAP report to PaNAGAIC, http://www.ifae.es/henap/
\bibitem{SolarNuTalks}
  See review talks by J. N. Bahcall, G. Bellini, V. Gavrin, A. Hallin, T. Kirsten, K.
  Lande, S. Sch\"{o}nert \& M. Smy, these proceedings.
\bibitem{fly}
  Bird, D. J. {\it et al.} 1994, Astrophys. J. {\bf 424}, 491.
\bibitem{agasa}
  Hayashida N. {\it et al.} 1999, Astrophys. J. {\bf 522}, 225
  and astro-ph/0008102.
\bibitem{composition}
  Dawson, B. R., Meyhandan, R., and Simpson, K. M., 1998, Astropart. Phys. {\bf 9},
  331; Abu-Zayyad, T. et al. 2001, ApJ {\bf 557}, 686.
\bibitem{Watson91}
  Watson, A. A. 1991, Nuc. Phys. B (Proc. Suppl.) {\bf 22}, 116.
\bibitem{anisotropy}
  Bird, D. J. {\it et al.} 1998, Astrophys. J. {\bf 511}, 739;
  Hayashida N. {\it et al.} 1999, Astropart. Phys. {\bf 10}, 303.
\bibitem{alternatives}
  See review talks by J. Learned \& A. Letessier-Selvon, these
  proceedings.
\bibitem{SiglNaganoWatson}
  Bhattacharjee, P. \& Sigl, G. 2000, Physics Reports, 327, 109;
  Nagano, M. \& Watson, A. A. 2000, Rev. Mod. Phys. 72, 689.
\bibitem{BnW02}
  Bahcall, J. N. \& Waxman, E. 2002, submitted to Phys. Lett. B, hep-ph/0206217.
\bibitem{HiRes}
  Abu-Zayyad, T., et al., 2002, submitted PRL, astro-ph/0208243.
\bibitem{yakutsk}
  N. N. Efimov {\it et al.}, in {\it Proceedings of the
  International Symposium on Astrophysical Aspects of the Most Energetic
  Cosmic-Rays}, edited by M. Nagano and F. Takahara (World Scientific,
  Singapore 1991), p. 20.
\bibitem{HP}
  Hinton, J., et al., Proc. Ultra High Energy Particles from Space,
  Aspen 2002 (http://hep.uchicago.edu/~jah/aspen/aspen2.html).
\bibitem{Yoshida95}
  Yoshida, S., {\it et al.}, Astropar. Phys. {\bf 3}, 151 (1995).
\bibitem{SFR}
  S. J. Lilly, O. Le Fevre, F. Hammer and D. Crampton,
  Astrophys. J.  {\bf 460}, L1 (1996);
  P. Madau, H. C. Ferguson,
  M. E. Dickinson, M. Giavalisco, C. C. Steidel and A. Fruchter,
  Mon. Not. Roy. Astron. Soc.  {\bf 283}, 1388 (1996).
\bibitem{W95a}
  Waxman, E. 1995, Astrophys. J. Lett. {\bf  452}, L1.
\bibitem{GZK}
  K. Greisen 1966, Phys. Rev. Lett. {\bf 16}, 748;
  Zatsepin, G. T. and Kuzmin, V. A. 1966, JETP {\bf 4}, 78.
\bibitem{clustering}
 Bahcall, J. N. and Waxman, E. 2000, Astrophys. J. {\bf 542}, 542.
\bibitem{WBbound}
  Waxman, E. and Bahcall, J. N. 1999, Phys. Rev. D {\bf 59},
  023002; Bahcall, J. N. \& Waxman, E. 2001, Phys. Rev. D {\bf 64},
  023002.
\bibitem{thick} Stecker, F., Done, C., Salamon, M., and Sommers,
  P. 1991, Phys. Rev. Lett. {\bf 66}, 2697,  erratum
  Phys. Rev. Lett. {\bf 69}, 2738 (1992); Berezinsky, V. S. and Dokuchaev,
  V. I. 2002, Nuc. Phys. B (Proceedings Supplements), {\bf 110}, 522.
\bibitem{exp-bounds}
  See review talks by G. Domogatsky \& D. Cowen, these proceedings.
\bibitem{AG_th_review}
  For updated review see M\'esz\'aros, P. 2002, ARA\&A 40, 137.
\bibitem{Blandford87}
  For review see, e.g., Blandford, R., \& Eichler, D. 1987, Phys. Rep. {\bf 154}, 1.
\bibitem{W95}
  Waxman, E. 1995, Phys. Rev. Lett. {\bf 75}, 386.
\bibitem{W01rev} Waxman, E. in
  {\it Physics \& Astrophysics of Ultra-High-Energy Cosmic Rays},
  Eds. M. Lemoine \& G. Sigl (Springer),
  Lecture Notes in Physics {\bf 576}, 122-154 (2001).
\bibitem{W02}
  Waxman, E. 2002, submitted to ApJL (astro-ph/0210638).
\bibitem{WnB-AG}
  Waxman, E., \& Bahcall, J. N. 2000, Ap. J. {\bf 541}, 707.
\bibitem{Schmidt01}
  Schmidt, M. 2001, Astrophys. J. {\bf 552}, 36.
\bibitem{Frail01}
  Frail, D. A. {\it et al.} 2001, Astrophys. J. Lett. {\bf 562}, L55.
\bibitem{WnB97}
  Waxman, E., \& Bahcall, J. N. 1997, Phys. Rev. Lett. {\bf 78}, 2292.
\bibitem{GSW01}
  Guetta, D., Spada M., \& Waxman, E. 2001, Ap. J. {\bf 559}, 101.
\bibitem{Dai00}
  Dai, Z. G., \& Lu, T. 2001, Ap. J. {\bf 551}, 249.
\bibitem{MnW01}
  M\'esz\'aros, P., \& Waxman, E. 2001, Phys. Rev. Lett. {\bf 87}, 1002
\bibitem{Mirabel99}
  Mirabel, I. F. \& Rodriguez, L. F. 1999, ARA\&A, {\bf 37}, 409.
\bibitem{LnW01}
  Levinson, A. \& Waxman, E. 2001, Phys. Rev. Lett. 87, 171101.
\bibitem{diStefano02}
  Distefano, C., Guetta, D., Waxman, E., Levinson, A. 2002, ApJ 575, 378

\end{thebibliography}
\end{document}